\documentstyle[prl,aps,preprint]{revtex}

\def\beq{\begin{equation}}\def\eeq{\end{equation}}
\def\bea{\begin{eqnarray}}\def\eea{\end{eqnarray}}
\def\a{\alpha} \def\b{\beta} \def\g{\gamma} 
\def\d{\delta}  \def\ee{\epsilon}
 \def\th{\theta}  
  \def\L{\Lambda} 
  \def\p{\pi}  
  \def\t{\tau} 
 \def\f{\phi} \def\F{\Phi} 
\def\Ps{\Psi}  \def\O{\Omega} 
\def\pa{\partial}  
 \def\IR{I\!\!R}

\begin{document}
\setcounter{page}{0}
\def\footnoterule{\kern-3pt \hrule width\hsize \kern3pt}
\tighten
\title{On the Wavefunctional for Two Heavy Color Sources
in Yang-Mills Theory\thanks
{This work is supported in part by funds provided by the U.S.
Department of Energy (D.O.E.) under cooperative
research agreement \#DF-FC02-94ER40818, and by NSF Grant
PHY-92-06867.}}

\author{Peter E. Haagensen and Kenneth Johnson}

\address{{~}\\
Center for Theoretical Physics \\
Laboratory for Nuclear Science \\
and Department of Physics \\
Massachusetts Institute of Technology \\
Cambridge, Massachusetts 02139 \\
{\rm e-mail:} {\tt haagense@ctp.mit.edu,
knjhnsn@mitlns.mit.edu}\\
{~}}

\date{MIT-CTP-2614, ~ hep-th/9702204, ~ February 1997}
\maketitle

\thispagestyle{empty}

\begin{abstract}

In an abelian gauge theory, the Coulombic potential between
two static charges is obtained most directly when a correct separation
between gauge-invariant and gauge degrees of freedom is made. This
motivates a similar separation in the nonabelian theory. When a careful
identification
of the Hilbert space is made, along with the proper analyticity
requirements,
it is then possible to find the appropriate wavefunctionals
describing
heavy color sources in the theory. This treatment is consistent
with, and realizes in a simple way the center
symmetry $Z_N$ of $SU(N)$ gauge theories.

\end{abstract}

\vspace*{\fill}

\pacs{xxxxxx}

\section{introduction}

Anyone who has given a certain amount of thought to different
treatments of quantum electrodynamics comes inevitably to the
conclusion that the reason it is a rather straightforward
theory to work out is, in
one guise or another, always the same: at some point, one
needs to clearly separate what are physical degrees of freedom
from what are gauge degrees of freedom, and this separation is
absolutely clear in an abelian gauge theory. Unfortunately,
no such luck persists with nonabelian gauge theories.
The nonlinear nature of the theory obscures considerably this
separation of physical from gauge components. One might nevertheless
wonder whether a similar systematic separation of degrees of freedom
is possible in the nonabelian theory as well and, if so, whether it
may shed some light on different aspects of the theory in any way.

In previous work \cite{us}, we have given a description
of the Hilbert space of Yang-Mills theory in terms of
gauge invariant variables which, as we show in what follows,
does achieve this objective without tremendous
technical effort. The specific issue we address here
is the form state functionals of the theory should take in the
presence of two static color sources. Firstly, it is of course
the nonzero charge sectors that exploit this separation of degrees
of freedom. Moreover, through the static charge-anti-charge state
one may also study the confining potentials generated by
the theory. The central points guiding the search
for these wavefunctionals are the above mentioned separation of
degrees of freedom, and a proper identification of the Hilbert space of
the theory. We take the view here that the only two requirements we 
are allowed to make are that wavefunctionals
satisfy Gauss' law and that they be analytic functions in the 
Hilbert space of the theory. We find {\it a posteriori}
that wavefunctionals for two static color sources satisfying these
two requirements necessarily contain a dependence on some curve
connecting the two charges if they belong to a fundamental 
representation, a behavior that is markedly different from both
the abelian case and the case of nonabelian adjoint sources. 
The reason this behavior comes about lies in the correct
identification of the Hilbert space in the nonabelian
theory, which in turn comes from the observation
that a pure $SU(N)$ gauge theory is in reality only an $SU(N)/Z_N$
gauge theory
because of the well-known fact that the gluons, which belong to
the adjoint representation, are insensitive to the center
transformations, $Z_N$. Specifically
for $SU(2)/Z_2=SO(3)$, the description of
the Hilbert space becomes particularly clear, through the use of
Euler angles and variables which simply rotate under gauge
transformations, and this is the case we will be concerned
with here, although the extension to $SU(N\! >\! 2)$ should not
present any conceptual difficulties.
Analyticity and periodicity requirements in the $SO(3)$ domain
will naturally
lead to a simple and explicit realization of the center symmetry,
through 't Hooft's operator algebra \cite{thooft}. At the same time,
Wilson loops are not gauge invariant under 
transformations generating vortices through them, and this fact would
be entirely missed if the gauge group were considered to be $SU(2)$
rather than $SO(3)$, unless the nontrivial $Z_2$ behavior were 
inserted by hand, {\it e.g.}, through boundary conditions \cite{thooftb}.

We begin by treating the abelian case, since this will not only
motivate the correct separation of degrees of freedom in the
nonabelian theory, but will also indicate some cautionary remarks
concerning the physical interpretation commonly ascribed to Wilson
loops, both in the abelian and the nonabelian cases.
The standard procedure to obtain the 
potential for two static charges in the abelian theory is to calculate the
vacuum expectation value of the Wilson loop \cite{kogut}
\beq\label{abelwilson}
< e^{i\oint dx^\mu A_\mu(x)}>\ \propto\ e^{{-1\over 8\p^2}
\oint\oint{dx^\mu dy^\mu\over (x-y)^2}}\ .\eeq
However, extracting a Coulomb law from Eq.~(\ref{abelwilson}) requires
some nontrivial steps. The integral contains  
divergences, and thus if one computes it for, say,
a rectangular loop of sides $L$ and $T$, the exact result is
a function of $\L L$ and $\L T$, symmetric under $L\leftrightarrow
T$ interchange, where $\L$ is a momentum cutoff. The double
integral then yields:
\beq\label{dblint}
c_1 \L (L+T) + c_2 (L/T+T/L)+ ...\ \eeq
(the exact result will be presented below). The Coulomb law is
buried in Eqs.~(\ref{abelwilson},\ref{dblint}) according to well-known
arguments: one must first interpret the expectation value to mean the
amplitude for creation of a charge-anti-charge pair, separated by a 
distance $L$,
out of the vacuum during some finite time $T$. Then,
one must argue away the first term as a self-energy contribution,
and let $T$ become large in order to isolate the behavior
${\rm exp}(-iT/L)$. Some further argumentation then allows one to 
identify the $1/L$ in this latter expression as the potential between
the charges running in the loop.
Altogether, we believe this entire procedure to be a bit too much 
calculation and argumentation. In a Hamiltonian
formalism, the same result can be seen to follow in a much simpler fashion.
We will also show, however, that unless one makes a judicious separation
of gauge and gauge-invariant variables, one may well erroneously
``derive" confinement in this theory. The lessons learned
in this exercise will then serve as a guide in treating the nonabelian 
theory. In particular, we will show that the form wavefunctionals
should take is not the one directly connected to Wilson loops, and
sometimes indicated in the literature.

\section{(No) Confinement in the Abelian Theory}

Here and in what follows, we consider the Hamiltonian
quantization of gauge theories in ``temporal" gauge
(in the abelian theory: $A_0=0$), and in the
Schr\"odinger representation. States of the theory
are wavefunctionals $\Phi [A]$ which
are eigenfunctions of the Hamiltonian, with the further
requirement that they satisfy Gauss' law, which in the
abelian theory reads
\beq\label{gauss}
\pa_k\ {\delta\F [A]\over\d A_k(x)}=0\ .\eeq

If two static charges, of opposite unit charge,
for definiteness, are placed at points $x_0$ and
$x_1$, states of the theory will be modified, and will
in particular solve a modified Gauss law:
\beq\label{gaussinhom}
\pa_k\ {\delta\Psi [A]\over\d A_k(x)}=-i(\d^{(3)}(x-x_1)-
\d^{(3)}(x-x_0))\Psi [A]\ .\eeq
If a functional $\F [A]$ satisfies Eq.~(\ref{gauss}), then
the functional
\beq\label{psi}
\Ps [A]=e^{i\int_{x_0}^{x_1}dx^m A_m(x)}\F[A]\eeq
will satisfy Eq.~(\ref{gaussinhom}). In the above, the line
integral is taken along some curve $C$ connecting $x_0$ and
$x_1$, and for simplicity we do not indicate explicitly
the dependence of $\Ps$ on $x_0$, $x_1$ or $C$. The resemblance
between this functional and the Wilson loop operator does suggest
the above to be the natural candidate for the state with opposite charges
at $x_0$ and $x_1$, if $\F$ is the vacuum state. The electric
field expectation on $\Ps$ is obtained from
\beq\label{electric}
E^i(x)\Ps [A]={1\over i}\ {\d\over\d A_i(x)}\Ps [A]=
\int_{x_0}^{x_1}dy^i \d^{(3)}(x-y)\ \Ps [A]-i
e^{i\int_{x_0}^{x_1}dy^m A_m(y)}{\d\F[A]\over\d A_i(x)}\ .\eeq
If we assume $\F$ is real -- we might take, for instance, $\F$ to 
be indeed the vacuum state -- there is no cross term in squaring
this expression and we are led to the following electric energy expectation:
\bea\label{elenergy}
\int d^3\!x <\! E^2(x)\! >_\Ps &=&
\int d^3\!x <\! E^2(x)\! >_\F +\int_{x_0}^{x_1}dy^m
\int_{x_0}^{x_1}dz^m \d^{(3)}(y-z)\\
&=&\int d^3\!x <\! E^2(x)\! >_\F +2L\d^{(2)}(0)\nonumber\ ,\eea
where $L$ is the length of the curve $C$, and
$\d^{(2)}(0)$ is the inverse thickness of the curve, to
be made finite by some suitable renormalization. This
is minimized for $C$ a straight line between $x_0$ and
$x_1$, and shows an electric energy growing linearly
with the separation of the charges. In other words,
confinement of charge as the term is usually understood.

What have we done wrong? We have simply chosen a very bad
functional: taking a real $\F$ has eliminated the cross
term in the (squared) electric energy, and has given a highly
divergent contribution. It is not that the derivation is
wrong, but there is a much better wavefunctional
around, leading to the exact solution, and a smaller electric energy. 
We find it
by separating the gauge variables from the gauge invariant variables:
\beq\label{separate}
A_i(x)=A_i^T(x)+\pa_i\L (x)\ ,\eeq
with $\pa_iA_i^T(x)=0$, and
\beq\label{lambda}
\L (x)=-{1\over 4\p}\int d^3\!y \ {\pa_iA_i(y)\over |x-y|}\ .\eeq
Inserted into Eq.~(\ref{psi}), it leads to
\bea\label{psinew}
\Ps [A]&=&e^{i(\L (x_1)-\L (x_0))}\left[
e^{i\int_{x_0}^{x_1}dy^m A^T_m(y)}\F[A]\right]\\
&\equiv&e^{i(\L (x_1)-\L (x_0))}\tilde{\F}[A]\ ,\nonumber\eea
where $\tilde{\F}[A]$ is gauge invariant and thus
still solves the source-free Gauss law, Eq.~(\ref{gauss}). 
The electric field now reads
\bea\label{elnew}
{1\over i}\ {\d\over\d A_i(x)}\Ps [A]&=&\left( {\d\L (x_1)\over\d
A_i(x)}-{\d\L (x_0)\over\d A_i(x)}\right)\Ps [A] -ie^{i(\L (x_1)-\L
(x_0))}{\d\tilde{\F}[A]\over\d A_i(x)}\\
&=&\pa_i\left( {1\over 4\p|x-x_1|}-{1\over 4\p|x-x_0|}\right)\Ps
[A]-ie^{i(\L (x_1)-\L (x_0))}{\d\tilde{\F}[A]\over\d
A_i(x)}\nonumber\ .\eea
In squaring this expression to obtain the electric energy, the
cross term vanishes without any further
assumptions since $\tilde{\F}$ is gauge invariant and thus satisfies 
Eq.~(\ref{gauss}). The electric energy then reads:
\beq\label{elennew}
\int d^3\!x <\! E^2(x)\! >_\Ps =\int d^3\!x <\! E^2(x)\! 
>_{\tilde{\F}} -2\ {1\over 4\p |x_0-x_1|}+{\rm s.e.}\ ,\eeq
where ``s.e." means the self-energy terms for both sources. Now
we see the correct result, no confinement, just the Coulomb interaction.
Whereas previously our choice of prefactor in Eq.~(\ref{psi}) led to the
singular quantity $\d A_i(x)/\d A_j(y)\propto \d^{(3)}
(x-y)$, the second choice
led on the other hand to the smoother quantity $\d\L (x)/\d
A_i(x)$. A seemingly ``natural" choice of $\F$ ({\it i.e.}, $\F$ 
the vacuum wavefunctional) then
did not allow for the cancellation of this extraneous divergence.

We thus learn that when we introduce a prefactor into
wavefunctionals in order to solve the Gauss law with sources, the
presence of the {\it full} vector potential in the line integral will lead
to an artificially divergent quantity; one should rather attempt to
introduce only the gauge part of the vector potential. In the
abelian case, the gauge part of the vector potential appears in such a
way as to effectively ``disconnect" the two end points where the charges
are located, without introducing any discontinuities
(cf. Eq.~(\ref{psinew})); we may suspect that in a correct treatment of the
nonabelian theory with fundamental sources
this should not happen, as in that case one does expect a
``line" to be present connecting the two sources.

It is possible in fact to derive the correct result in a simpler
way by directly solving the local form of Gauss' law, because it
makes no premature use of a path: using the appropriate separation of
gauge and physical variables, Eq.~(\ref{separate}), it follows that
\beq\label{ggg}
\pa_k\ {\delta\Psi\over\d A_k(x)}=-{\d\Psi\over\d \L (x)}\ ,\eeq
which in turn leads to, using Eq.~(\ref{gaussinhom}),
\beq\label{hhh}
{\d\Psi\over\d \L (x)}=i(\d^{(3)}(x-x_1)-
\d^{(3)}(x-x_0))\Psi\ ,\eeq
which immediately gives the correct solution:
\beq\label{corr}
\Ps [A]= e^{i(\L (x_1)-\L (x_0))} \F_0 [A^T]\ .\eeq
Further minimization in the gauge invariant sector tells us that
$\F_0$ is the vacuum wavefunctional. Incidentally, we note that 
this is not at all a new result: it can in fact be traced back to
Dirac \cite{dirac}.

The separation of gauge and 
gauge invariant variables has clearly indicated that it is the 
longitudinal degrees of freedom which are responsible for the 
potential between static charges, and once that separation is
made, Gauss' law by itself allows for the computation of the 
potential. With that in mind, one understands why the same result
is more awkward to extract from the Wilson loop, since that is
an expectation value of a gauge invariant operator in the vacuum
sector of the theory, which only depends on transverse variables.
We have furthermore learned that the wavefunctional which is
suggested by the Wilson loop, and which might 
seem at first more natural (Eq.~(\ref{psi})) in fact is incorrect,
essentially because it introduces all of the gauge field degrees of freedom,
transverse and longitudinal, propagating along the line connecting
the two charges. Given this result, however, one might now wonder:
what has conspired to allow the Wilson loop expectation to also
indicate the correct result, albeit with more effort? In order to answer
this question, we first observe that the wavefunctional in Eq.~(\ref{corr})
is an exact eigenfunction of the Hamiltonian operator
\beq\label{hamiltonian}
H={1\over2}\int d^3\! x \left( -{\d^2\over\d A_i(x)\d A_i(x)}+
B_i^2(x)\right)\ ,\eeq
with eigenvalue $v_0\L^4V+v_1\L+v_2/|x_0-x_1|$, where $v_i$ are constants, 
$\L$ is a momentum cutoff, $V$ is the volume of space, the first 
term is a vacuum zero-point energy, the second term is a self-energy, and the
last term is the Coulomb energy. Meanwhile, it is also simple to verify
explicitly that the Wilson line, Eq.~(\ref{psi}) (with $\F$ the vacuum
functional), is {\it not} an eigenstate of $H$. In order to make contact
between the Wilson loop and our calculations, we may choose to calculate 
the former in the gauge $A_0=0$, for a loop tracing a rectangle, with two 
fixed-time
sides of length $L$ and two sides of length $T$ along the time direction.
Due to the gauge choice, the expectation value only receives contributions
from the spacelike sides. Furthermore, each of these sides corresponds
to a state given by a Wilson line acting on the vacuum state, Eq.~(\ref{psi}).
Since one is at time $T$ and the other at time $0$, we may insert the
operator $\exp -iTH$ in the middle in order to bring these states to the same
time, and this operator can be further decomposed into its matrix elements
on energy eigenstates. Then, the lowest nonvanishing overlap will give the
leading exponential behavior of the loop. Because the Wilson line belongs to
the sector with opposite charges at $x_0$ and $x_1$, the lowest
Hamiltonian eigenstate with a nonvanishing overlap
will be precisely the eigenstate given by Eq.~(\ref{corr}), 
so that the correct
Coulomb behavior is obtained. If the Wilson line only had
components in the higher modes within this two-charge sector, the expectation
value of the Wilson loop would be even further removed from the Coulomb law.

In order to explicitly confirm the relevance of this overlap, we have 
performed the exact computation of the Wilson loop expectation value for a 
rectangle of sides $L_1$ and $L_2$ along two spatial directions.
It is a lengthy computation with, however, a rather compact result: we find
\beq\label{exactloopone} 
{1\over 8\p^2}\oint\oint{dx^\mu dy^\mu\over (x-y)^2}={1\over4\p^{3/2}}
{1\over\sqrt{s_0}}(L_1+L_2)-{1\over4\p}\left( {L_1\over L_2}+{L_2\over L_1}
\right)\eeq
\vspace{-.1cm}
$$ -{1\over2\p^2}\left( -\ln \left[ 4s_0\left( {1\over L_1^2}+
{1\over L_2^2}\right)\right] +\g +2\right)
+{1\over2\p^2}\left( {L_1\over L_2}{\rm tan}^{-1}{L_2\over L_1}+
{L_2\over L_1}{\rm tan}^{-1}{L_1\over L_2}\right)\nonumber\ ,$$
where $s_0\to 0$ is a cutoff parameter (of dimension $L^2$) used to regulate
the propagator in a Schwinger parametrization, and $\g$ is the Euler
constant (we have also ommitted two terms vanishing in the limit $s_0\to 0$).
To orient the loop along the 
time direction and extract the Coulomb law, we take $L_1>>L_2$ and then
$L_1\to -iT$, and are left with
\beq\label{larget}
\exp\left\{ -iT\left( {1\over4\p^{3/2}}
{1\over\sqrt{s_0}}-{1\over4\p L_2}\right)
-{1\over2\p^2}\left[\ln \left( {L_2^2\over 4s_0}\right) +\g +1\right]
\right\}\ .\eeq
The first term is the self-energy contribution, the second one is the Coulomb
law, and the last term is the (modulus squared) overlap of the Wilson line
with the correct eigenstate, Eq.~(\ref{corr}). As $L_2$ is made larger,
this overlap becomes smaller, which is expected, as for larger 
separations the Wilson line becomes more and more energetic, and thus
further removed from the true charge-anti-charge state. Furthermore, we also 
see that, as the cutoff is removed, this overlap becomes vanishing.
One may understand this as a renormalization of the composite operator
corresponding to the Wilson line, but however one wants to interpret it,
it does mean that the overlap does indeed vanish in the limit
$s_0\to 0$.

As long as the Wilson line has a nonzero overlap
with the correct wavefunctional for two charges, the Wilson loop
expectation value should in the end
lead to the correct potential between static charges. If, for instance, on a 
lattice calculation one is not forced to 
take the analogous limit $s_0\to 0$ in 
order to extract the appropriate 
scaling behaviors, then such a nonzero overlap
may well be achieved. At the same
time, however, we have seen that there are {\it caveats} to the full 
computation, even in the simpler abelian case where an exact result can be 
found, and in general Wilson loops
should not be taken as any direct indication of what the correct 
wavefunctional is for a pair of static charges. Some of the literature on
this subject may be a bit misleading, as sometimes it implicitly appears
to assume the analogous form of Eq.~(\ref{psi}) for the wavefunctional in
the nonabelian case, and in some instances such a wavefunctional is explicitly
written down \cite{wilson}.

\section{the nonabelian theory}

Much of what we do here will be valid for general $SU(N)$ gauge
group, but in order to keep the discussion sufficiently
simple and explicit, we will treat the $SU(2)$ theory.
In the presence of two static color sources at $x_0$ and $x_1$,
state wavefunctionals take the form
\beq\label{state}
\Ps_{\a\b}[A]\ ,\eeq
where $\a$ and $\b$ are indices in the representation of the
sources at $x_1$ and $x_0$, respectively. Under finite gauge
transformations, these states transform as
\beq\label{statetrsf}
\Ps_{\a\b}[A]\longrightarrow\O_U^{-1}\Ps_{\a\b}[A]=\Ps_{\a\b}[{}^
U\! A]=U_{\a\a '}(x_1)U_{\b\b '}^*(x_0)\Ps_{\a '\b '}[A]\ .\eeq
For notational simplicity, we have written the same ``$U$"
everywhere, but of course it should be kept in mind that $A$ transforms in
the adjoint, while the $U$'s at $x_0$ and $x_1$ belong to whatever
representation the
sources are in. Here we are mostly interested in the case in which both
sources are either in the fundamental or in the adjoint.
A physical wavefunctional should transform as
the above under gauge transformations,
and the infinitesimal statement of such a transformation is
Gauss' law in the presence of sources:
\beq\label{gaussnonab}
\left( {\cal G}^a(x)+i\d^{(3)}(x-x_1)\L^a-i\d^{(3)}(x-x_0)\L^{a*}
\right)_{\a\b ,\a '\b '}\Ps_{\a '\b '}=0\ ,\eeq
where $\L^a$ is an $SU(2)$ generator, either in the fundamental
or in the adjoint, and
\beq\label{ga}
{\cal G}^a(x)=D_i^{ab}\ {\d\over\d A_i^b(x)}\ ,\eeq
with $D_i^{ab}=\pa_i\d^{ab}+\ee^{acb}A_i^c$,
is the infinitesimal generator of gauge transformations on functionals
of $A$ (in the above, unit matrices in $\a\a ',\b\b '$-space are
also understood in the appropriate places).

As previously, one solution of this constraint is gotten by
attaching a Wilson line prefactor to a gauge invariant functional
$\F$:
\beq\label{wline}
\Ps_{\a\b} [A]=P\left(e^{i\int_{x_0}^{x_1}dx^m
A_m(x)}\right)_{\a\b}\F[A]
\ ,\eeq
where $P$ stands for the usual path ordering prescription, and
$A_m(x)=A_m^a(x)\L^a$ is a matrix in the algebra of $SU(2)$ in
the representation of the sources. This functional transforms
as in Eq.~(\ref{statetrsf}), and consequently satisfies Gauss'
law in the presence of sources.
Furthermore, it is a single-valued, analytic function of the
vector potential in the $SO(3)$ domain of the variables,
irrespective of whether $\L^a$ is a single-valued representation or not.
However, like in the abelian case, it leads to an extraneous
divergence when
calculating the electric field energy, due to the singular term
$\d A_i(x)/\d A_j(y)\propto \d^{(3)}(x-y)$, and we must discard it.

What we should seek, rather,
is the equivalent of the abelian term $\d\L (x)/\d A_i(x)$ in the
electric energy,
while at the same time preserving the transformation property
Eq.~(\ref{statetrsf}) and single-valuedness in the
$SO(3)$ domain of the variables.

The abelian example suggests that we attempt to solve Gauss' law
in local form, since it led most simply to the correct result in
that case. In order to do so, we have found it most convenient to utilize a
set of gauge covariant variables that we have described in previous work
\cite{us}.
These new variables, $u_i^a$, replace the vector potential
through the following transformation:
\beq\label{edu}\ee^{ijk}D_ju^a_k\equiv
\ee^{ijk}(\pa_ju^a_k+\ee^{abc}A^b_j u^c_k)\ =\ 0\ .\eeq
Unlike the vector potential, which transforms inhomogeneously,
these new variables simply rotate under gauge transformations:
\beq\label{transfu}
\O_U^{-1}u^a_i(x)\O_U\ =\ U^{ab}(x)u^b_i(x)\ ,\eeq
with $U^{ab}$ a $3\times 3$ real orthogonal matrix.
From these, furthermore, gauge invariant variables can easily be
obtained by contraction in color indices: $g_{ij}=u^a_iu^a_j$. The nine
degrees of freedom of $A_i^a$ are thus clearly split between six gauge
invariant degrees of freedom in $g_{ij}$ and three $SO(3)$ 
(rather than $SU(2)$) gauge angles. For our considerations
here, it will suffice to consider the variables $u_i^a$. Taken as
three vectors in color space, for $i=1,2,3$, they uniquely define a
tetrahedron at each point of $\IR^3$,
with vertices at the origin and at the ends of the vectors.
Variations in the gauge invariant variables change the shape or the size
of the tetrahedron, while gauge variations simply rotate it. In
terms of the new variables, the Gauss law generator reads
\beq\label{gaussu}
{\cal G}^a(x)=\ee^{abc}u_i^b(x){\d\over\d u_i^c(x)}\ .\eeq
To satisfy Eq.~(\ref{gaussnonab}) in the simplest possible way,
one may take a wavefunctional
\beq\label{gausssol}
\Ps_{\a\b}[u_i^a(x);u_i^a(x_0),u_i^a(x_1)]\ ,\eeq
which is a functional of $u_i^a(x)$ throughout space, but a
regular function of $u_i^a(x_0)$ and $u_i^a(x_1)$. Then,
\beq\label{ddu}
{\d\Ps\over\d u_i^a(x)}=\left. {\d\Ps\over\d
u_i^a(x)}\right|_{u_0,u_1}+
\d^{(3)}(x-x_0){\pa\Ps\over\pa u_i^a(x_0)}+
\d^{(3)}(x-x_1){\pa\Ps\over\pa u_i^a(x_1)}\ ,\eeq
where the functional derivative on the r.h.s.~is taken with
$u_i^a(x_0)$ and $u_i^a(x_1)$ fixed. Gauss' law becomes
\beq\label{gaussone}
\left(\ee^{abc}u_i^b(x_0)\ {\pa\over\pa u_i^c(x_0)}-
i\L^{a*}\right)\Ps_{\a\b}=0\ \eeq
at $x_0$,
\beq\label{gausstwo}
\left(\ee^{abc}u_i^b(x_1)\ {\pa\over\pa u_i^c(x_1)}+
i\L^a\right)\Ps_{\a\b}=0\ \eeq
at $x_1$, and
\beq\label{gaussthree}
\ee^{abc}u_i^b(x)\left.
{\d\Ps_{\a\b}\over\d u_i^a(x)}\right|_{u_0,u_1}=0\ .\eeq
If we associate Euler angles $(\a ,\b, \g )_0$ and
$(\a ,\b, \g )_1$ to each of the tetrahedra
$u_i^a(x_0)$ and $u_i^a(x_1)$, the rotation generators in
Eqs.~(\ref{gaussone},\ref{gausstwo}) become
the angular momentum operators
\bea\label{angmom}
J_3&=&-i{\pa\over\pa\a}\\
J_{\pm}&=&e^{\pm i\a}\left[\pm{\pa\over\pa\b}+{1\over\sin\b}
\left( {1\over i}{\pa\over\pa\g}-\cos\b{1\over i}{\pa\over\pa\a}
\right)\right]\nonumber\ ,\eea
at $x_0$ and $x_1$, with $J_{\pm}=J_1\pm iJ_2$.

The solution to Eqs.~(\ref{gaussone},\ref{gausstwo}) is
well-known from
the representation theory of $SU(2)$. If the $\L^a$ belong to the
spin-$j$ representation, it is given by the Wigner ${\cal D}^{(j)}$
functions, which satisfy:
\beq\label{wignerd}
J^a{\cal D}^{(j)}_{mm'}=-\L^a_{mn}{\cal D}^{(j)}_{nm'}\ ,\eeq
where $m,m',n$ span integer values from $-j$ to $j$. In the case
of interest for us, $j=1/2$, $\L^a$ are the Pauli matrices
$\t^a/2$, and the complete solution to
Eqs.~(\ref{gaussone}-\ref{gaussthree}) is given by:
\beq\label{solution}
\Ps_{\a\b}={\cal D}^{(1/2)}_{\a m}((\a ,\b ,\g )_1)
{\cal D}^{(1/2)*}_{\b n}((\a ,\b ,\g )_0)\F^{mn}[g_{ij}]\ ,\eeq
where $\F^{mn}[g_{ij}]$ are four arbitrary
gauge-invariant
functionals (and thus functionals of $g_{ij}$ only), with no
particular symmetry associated to the indices $m,n$.
These indices $m,n$ live in the same (spin-$1/2$) space as $\a
,\b$,
however, they are denoted in this way precisely to indicate the
lack of any
symmetry covariance associated to them, as far as Gauss' law is
concerned.
The ${\cal D}^{(1/2)}$ matrices appearing above are given by:
\beq\label{d}
{\cal D}^{(1/2)}_{mn}(\a ,\b ,\g )=
\left( \matrix{e^{-{i\over2}(\a +\g )}\cos \b /2 & -
e^{-{i\over2}(\a -\g )}\sin \b /2\cr
e^{{i\over2}(\a -\g )}\sin \b /2 &
e^{{i\over2}(\a +\g )}\cos \b /2  }\right)\ .\eeq

One more time, it seems we have been able to solve Gauss' law in
the presence of two sources. Furthermore, it also seems we have
eliminated the problem
of introducing ``too much" of the electric field propagating
between the two points. In fact, in this solution the points
$x_0$ and $x_1$ are entirely disconnected: to a given 
configuration $A_i^a(x)$ is associated a configuration of
Euler angles at each point in space, $(\a (x),\b (x),\g (x))$,
and to build Eq.~(\ref{solution}), one simply reads off the 
values of these angles at $x_0$ and $x_1$.

But now it becomes clear that we have eliminated the previous
problem at the cost of introducing another type of extraneous
divergence, and this solution will also have to be discarded.
The problem lies in the fact that the Euler angle variables
appearing in the angular
momentum operators, Eq.~(\ref{angmom}), originate in an
integer-spin representation of
$SU(2)$, and thus take values in the $SO(3)$ range $0\! \le\! \a\! <\! 2\p,
0\!\le\!\b\!\le\!\p ,0\!\le\!\g\! <\! 2\p$ (after all,
they represent rotations of a tetrahedron in $\IR^3$).
It is in this $SO(3)$ range that
Eqs.~(\ref{gaussone},\ref{gausstwo}) must be solved. However, by
the time we solve them
for $j=1/2$ sources, Eqs.~(\ref{solution},\ref{d}), we have
had to resort to the
fundamental representation of $SU(2)$, which is a double-valued
representation of $SO(3)$, with range $0\!\le\!\a\! <\! 4\p, 0\!\le\!
\b\!\le\!\p ,0\!\le\!\g\! <\! 2\p $. This means that when $\a$ (at either
$x_0$ or $x_1$) is rotated by $2\p$, which is no change at all
in the $SO(3)$ degrees of freedom, this solution changes sign.
From the $SO(3)$ viewpoint, the ${\cal D}^{1/2}$ matrices
undergo a discontinuous jump between $\a\! =\! 2\p$ and $\a\! =\! 0$.
Of course, this has happened because local solutions such as
Eq.~(\ref{solution}) cannot be expected to automatically discern
between $SU(2)$ and $SO(3)$ degrees of freedom, a question which 
can only be decided at a global level. 
We are thus led to discard Eq.~(\ref{solution}) as an appropriate
solution to the Gauss law constraint with fundamental sources due to the
double-valued and non-analytic nature of this solution.
We also see, on the other hand, that the situation is entirely
different with adjoint sources, where a solution would be
Eq.~(\ref{solution}) with the substitution
${\cal D}^{(1/2)}\rightarrow {\cal D}^{(1)}$ or, equivalently,
\beq\label{soladj}
\Ps^{ab}\ =\ u_i^a(x_1)u_j^b(x_0)\F^{ij}[g]\ .\eeq
It is trivial to see that this satisfies 
Eqs.~(\ref{gaussone},\ref{gausstwo}) with $\L^a$ in the adjoint
and, in this case, $\Ps$ would in fact be
a single-valued function of the Euler angles in the $SO(3)$
domain. We would furthermore be able to conclude that in
this case no line appears connecting the points $x_0$ and
$x_1$, much like in the abelian case, and this seems to indicate there
should not be a confining potential between these sources. At the same
time, there are clear indications from lattice calculations \cite{ambjorn}
that for a certain range of distances between the confinement and color
screening scales, adjoint and higher representation Wilson loops do 
experience an area law behavior with a string tension proportional to the
quadratic Casimir of the representation. While we are not questioning these
results, what our results point out, however, is that one should exercise
much caution in directly translating these Wilson loop scalings into the 
true potential experienced by static quarks.

In attempting to find an appropriate solution to
Eqs.~(\ref{gaussone}-\ref{gaussthree}) with sources in the
fundamental, we are guided by the following prerequisites: {\it i)} we
should not introduce all of the vector potential in a Wilson line prefactor,
but rather only the ``gauge part" of it, and
{\it ii)} we should seek a functional which is an analytic
function of the gauge variables in the $SO(3)$ domain, so that
the wavefunctional belongs to the correct Hilbert space. In separating the
gauge part of the vector potential, the $u_i^a$ variables are again very
useful. For, if we imagine that to each possible tetrahedron in
space we associate a unique triplet of vectors $v_i^a(x)$, then the
$u_i^a$ variables will in general be $SO(3)$ rotations of these,
\beq\label{uuv} u_i^a(x)=U^{ab}(x)v_i^b(x)\ ,\eeq
under gauge transformations. One must of course find a systematic
way to assign a unique $v_i^a$ to each tetrahedron, and
once that is done, the purely gauge degrees of freedom become
neatly separated in Eq.~(\ref{uuv}). Although it is not of central
importance for our
purposes here, we briefly describe how tetrahedra are associated
to $v_i^a$:
given a tetrahedron, we align its longest edge with the $z$-axis
with one
vertex at the origin; we then align the longest edge connected to
this first edge to lie on the
left $xz$-plane ($x<0$), also with one vertex at the origin. Then
the last
vertex will lie somewhere in $\IR^3$. The $v_i^a$ are then given
by the three
vectors describing the edges emanating from the origin, and some
thought
shows that the integration over gauge-inequivalent configurations
corresponds to
\beq\label{intregion}
\int_0^\infty dr_1\int_0^{r_1} dr_2\int_0^{r_2} dr_3
\int_0^\p d\th_{12}\int_0^\p d\th_{13}\int_0^{2\p} d\f \ ,\eeq
where $r_i=\sqrt{v_i^av_i^a}=\sqrt{g_{ii}},
\th_{ij}=v_i^av_j^a/r_ir_j=g_{ij}/\sqrt{g_{ii}g_{jj}}$ (no sum in
$i$ or $j$ in these expressions), and $\f$ is the
azimuthal angle of $v_3^a$ (the vector not lying in the
$xz$-plane).

This separation of gauge- and gauge-invariant variations is
sufficient to build the wavefunctional
we seek: it will be given by associating the $U^{ab}$,
corresponding to a
particular $u_i^a$, to a pure gauge vector potential
\beq\label{utoa}
U^{ab}\longrightarrow ({\cal A}_i)_{\a\b}(x)\equiv -{1\over2}
\ee^{abc}
U^{bb'}(x)\pa_iU^{cb'}(x)\left(\t^a/2\right)_{\a\b}\ .\eeq
This vector potential is then used in constructing the Wilson
line prefactor, leading to the candidate wavefunctional
\beq\label{candidate}
\Ps_{\a\b} =P\left(e^{i\int_{x_0}^{x_1}dx^m {\cal
A}_m(x)}\right)_{\a\b}
\F[g]\equiv F_{\a\b}\F[g]\ .\eeq
It is worthwhile to work out the gauge transformation 
properties of this wavefunctional
in order to clarify the issues of analyticity and center symmetry.
This is done by looking at the regularized form of its infinite product
representation:
\beq\label{infprod}
F^{\rm reg}=\prod_{n=1}^N (1+idx^i_n{\cal A}_i(x_n))\ ,\eeq
where we understand the curve connecting $x_0$ and $x_1$ to be
divided into a large number $N$ of segments at the points $\{ x_n\}$, 
and ${\cal A}_i={\cal A}_i^a\t^a/2$. The first term in this
product is
\beq\label{firstterm}
(1+idx^i_1{\cal A}_i(x_1))_{\a\b}=\d_{\a\b}-{i\over2}\ee^{abc}U^{bb'}(x_1)
dU^{cb'}(x_1)\left({\t^a\over2}\right)_{\a\b}\ ,\eeq
with $dU^{ab}(x_1)\equiv dx_1^i\pa_iU^{ab}(x_1)$. The standard commutation
and anticommutation relations of Pauli matrices then lead to
\beq\label{again}
(1+idx^i_1{\cal A}_i(x_1))_{\a\b}=\left({\t^a\over2}{\t^b\over2}\right)_{\a\b}
\left( {1\over3}\d^{ab}+U^{ab'}(x_1)U^{bb'}(x_2)\right)\ \eeq
to linear order, where $x_2=x_1-dx_1$. One then uses the standard
$2\! :\! 1$ map between $SU(2)$ and $SO(3)$,
\beq\label{sutoso}
U_{\a\a '}\left( {\t^a\over2}\right)_{\a '\b '}U_{\b '\b}^{-1}=
U^{ba}\left( {\t^b\over2}\right)_{\a\b}\ ,\eeq
and the ``spin-permutation" identity
\beq\label{tt}
\left({\t^a\over2}\right)_{\a\b}\left({\t^a\over2}\right)_{\a '\b
'}={1\over2}\left( \d_{\a\b '}\d_{\b\a '}-
{1\over2}\d_{\a\b}\d_{\a '\b '}\right)\ ,\eeq
to finally write
\beq\label{finally}
(1+idx^i_1{\cal A}_i(x_1))_{\a\b}=\left( U(x_1)U^\dagger (x_2)\right)_{\a\b}
[{1\over2}{\rm tr}_f (U(x_1)U^\dagger (x_2))]\ ,\eeq
with tr$_f$ being the trace in the fundamental. All $N$ terms in the product 
can now be put together to obtain
\beq\label{fullline}
F^{\rm reg}_{\a\b}
=(U(x_1)U^\dagger (x_0))_{\a\b}[{1\over2}{\rm tr}_f (U(x_1)U^\dagger (x_2))]
\ldots [{1\over2}{\rm tr}_f (U(x_N)U^\dagger (x_0))]\ .\eeq
In the continuum limit one might be tempted to simply drop the 
infinite trace term
since each trace involves two matrices which are very close together,
and to linear order differ by an element of the algebra, which is traceless.
This is almost correct, except for the fact that it endows the path
ordered exponential with the $SO(3)$ analyticity we have to require of
state wavefunctionals. That this is so can be seen as follows: we would 
like our candidate wavefunctional, Eq.~(\ref{candidate}), to be invariant
under the $SO(3)$-periodic change $\a\rightarrow\a +2\p$, independently at
$x_0$ and $x_1$. This is equivalent to making a singular gauge transformation
equal to the identity everywhere except for one of the points $x_0$ or $x_1$,
where it is equal to a rotation of $2\p$ around the $z$-axis. Then, either
$U(x_1)$ or $U(x_0)$ will change sign, but since $F^{\rm reg}$
is quadratic in the fundamental matrices at every point, it will remain
invariant. By simply dropping the trace term, this periodicity would be
broken. It is this analyticity in the $SO(3)$ domain that we must preserve
in taking the continuum limit of Eq.~(\ref{fullline}). Due to
continuity, furthermore, the trace factor allows for a {\it definition} of
the angles at $x_1$ starting from a given convention for angles at $x_0$.
Thus, if the angles at $x_0$ are taken to be in the $SO(3)$ domain, then,
in the continuum limit, the angles at $x_1$ will be given by
\beq\label{angles}
\a^C(x_1)=\a (x_0)+\int_C dx^i\pa_i\a (x), \eeq
and likewise for $\b^C (x_1)$ and $\g^C (x_1)$, where we write the 
superscript $C$ over the angles simply to differentiate them from the 
angles obtained by simply reading them off at the point $x_1$, as in the 
previous solution (thus, angles at $x_1$ have unlimited range). 
With this definition of angles
at $x_1$, we are now ready to drop the trace factors and take the continuum
limit for $F$:
\beq\label{F}
F_{\a\b}^C=(U^C(x_1)U^\dagger (x_0))_{\a\b}=
{\cal D}^{(1/2)}_{\a \g}((\a ,\b ,\g )^C_1)
{\cal D}^{(1/2)*}_{\b \g}((\a ,\b ,\g )_0)\ .\eeq
This represents Eq.~(\ref{candidate}) in a more explicit form. We note that
it satisfies Gauss' law while at the same time being
periodic in the $SO(3)$ domain. This is achieved through a residual
dependence on the line connecting the two charges.
Yet, it is very similar to the local
solution Eq.~(\ref{solution}), as it must be since both satisfy the same
local gauge invariance constraint. Furthermore, it emulates the abelian
case in the closest possible form, insofar as only gauge degrees of
freedom are introduced to ``communicate" between the two charges.

It is worthwhile noting that if one were to form a fundamental Wilson loop
in the fashion spelled out above, then it would transform with a $Z_2$
factor, $\pm 1$, under gauge tranformations, with the specific sign being 
determined by the number of vortex-like singularities in the gauge 
transformation that pierced a surface bounded by the loop. This is most 
clearly illustrated by a gauge transformation with a single infinite
vortex along the $z$-axis, which can be described as follows:
\beq\label{vortex}
\a (x,y,z)=\tan^{-1} {y\over x}\ ,\eeq
and $\b (x,y,z)=\g (x,y,z)=0$. Then, the Wilson loop will acquire a minus sign
under this gauge transformation if and only if it encircles the $z$-axis an
odd number of times. This shows how our analyticity requirement leads to a 
natural realization of 't Hooft's operator algebra \cite{thooft}
\beq\label{algthooft}
A(C)B(C')=e^{2\p in/N}B(C')A(C)\ ,\eeq
where (in the notation of \cite{thooft}) $A(C)$ is a Wilson loop around $C$
and $B(C')$ is the operator which creates a vortex at $C'$. For $SU(2)/Z_2$
gauge group, the phase on the r.h.s. above is equal to the ($\pm$) $Z_2$
factor described previously. If the gauge group were $SU(2)$ rather than
$SO(3)$ this $Z_2$ factor would naturally be absent, and Wilson loops would
be gauge invariant: in that case each point in space is uniquely associated
to an $SU(2)$ matrix and thus the matrices which transform the endpoints of
a Wilson line become one and the same when the endpoints meet, and the 
transformation vanishes in the trace. 

It should also be noted that in order to achieve an expression for
the wavefunctional of two sources which does not single out a preferred
curve, one must naturally consider wavefunctionals which are linear
combinations of the ones just described, going over different curves
connecting the two charges, with some weighting factor $w(C)$:
\beq\label{lincomb}
\Ps_{\a\b}=\sum_C w(C) F_{\a\b}^C \F [g]\ .\eeq
With such wavefunctionals at hand, one may of course then attempt 
to argue that confinement comes about due to random vortex fluctuations
in the vacuum, as in the standard $Z_N$ vortex condensation picture 
\cite{thooft}. At this point, however, we would rather refrain from such
speculations.

We finally observe that it is possible to exhibit the different
wavefunctionals considered here by means of a description which is
independent of the use of gauge covariant variables.
If we imagine one were to select a family of gauge potentials, $\hat{A}_i^a$, 
that crossed once and only once each gauge orbit, then any gauge potential
$A_i^a$ would be related to one of the $\hat{A}_i^a$ through a gauge 
transformation (of course, trying to achieve such a family of $\hat{A}_i^a$
through a gauge condition inevitably leads to Gribov ambiguities, but as 
long as we do not choose gauge conditions we are free to assume the 
existence of the $\hat{A}_i^a$ unambiguously). Then,
the generic solution to the Gauss law constraint should, strictly speaking, 
be
\beq\label{gensol}
\Ps_{\a\b}[A]=U_{\a m}(x_1)U_{\b n}^*(x_0)\F^{mn}[\hat{A}]\ ,\eeq
where $U_{\a\b}$ is the gauge transformation relating $A_i^a$ and
$\hat{A_i^a}$, and
$\F^{mn}[\hat{A}]$ is to be determined by a dynamical calculation
in the vacuum sector. The wavefunctional sometimes presented in the
literature, gotten by
acting with a fundamental Wilson line on the vacuum state is
\bea\label{wilsonline}
\Ps_{\a\b}[A]&=&P\left( e^{i\int A\cdot\t /2}\right)_{\a\b}\F_{\rm vac}
[\hat{A}]\\ &=&U_{\a m}(x_1)U_{\b n}^*(x_0)\left[
P\left( e^{i\int \hat{A}\cdot\t /2}\right)^{mn}\F_{\rm vac}
[\hat{A}]\right]\ .\eea
It clearly suffers from the same problem as the incorrect
wavefunctional in the abelian case, namely, the introduction of too
many gauge quanta propagating between the two sources. Our solution,
Eq.~(\ref{candidate}), would on the other hand, correspond to
\bea\label{oursolution}
\Ps_{\a\b}[A]&=
U_{\a m}(x_1)U_{\b n}^*(x_0)\left[ \d^{mn}\F_{\rm vac}
[\hat{A}]\right]\\
&=\left( U(x_1)U(x_0)^\dagger\right)_{\a \b}\F_{\rm vac}
[\hat{A}]\ .\eea
Of course, this description cannot be carried on much further, as we have 
no handle on how to describe the unique gauge slice $\hat{A}_i^a$.

\section{conclusions}

We have investigated here the form state functionals should take in the
presence of two static charges, both in abelian and in $SU(2)$ gauge
theory. Part of the motivation for such an investigation was the fact
that the functionals which are suggested by the form Wilson loops take,
are incorrect, and generate a confining static potential even where there
is no confinement. In the abelian case, we have shown how the static 
Coulomb potential arises in a considerably more direct fashion through
Schr\"odinger-picture methods than through 
Wilson loop expectation values. Another part of our motivation was 
to extend this treatment to the nonabelian theory, by exploring the 
gauge covariant description of the Hilbert space that we have proposed
in previous work. In the
nonabelian case, the use of gauge covariant variables has shown to be 
well-suited to the problem, as they allow for a direct solution of the
Gauss law constraint before the dynamical problem is tackled. Their use,
furthermore, together with an analyticity requirement, also led to a
simple and explicit realization of the center symmetry of the theory. 

The extension to gauge group $SU(3)$ should be essentially straightforward,
whereas the actual dynamical problem of determining the static potential
does present considerable technical difficulties. We have begun such an 
investigation, and hope to report on it in the near future.

\section{Acknowledgments}
It is a pleasure to thank our friends and colleagues Shailesh 
Chandrasekharan, Poul Henrik Damgaard, Andrei Matytsin and Uwe-Jens
Wiese for many pleasant and informative discussions.

\end{document}